**Astronomy job crisis**
*Astronomers outline changes for the academic system to promote a smooth transition for junior scientists from academia to industry.*

By Asantha Cooray, Alexandra Abate, Boris Häußler, Jonathan R. Trump and Christina C. Williams

This is the era of big astronomy, both from the ground and from space. A cursory reading of news articles alone is enough to see that the astronomers worldwide are making one discovery after another on planets, stars and galaxies that make up our Universe. The rapid growth in astronomical discoveries will continue to accelerate over the next decade with the James Webb Space Telescope, to be launched in 2018, joining Hubble, which just recently celebrated 25 years of successful scientific observations. On the ground, a new cohort of extremely large telescopes and new instruments for existing telescopes will complement the capabilities from space for deep and wide observations of the sky. Astronomy is also steadily moving away from projects led by a single investigator on shared national observatories to multi-institutional, multi-national large experiments that use dedicated telescopes. Such new efforts include multiple large collaborations involved with cosmological measurements that map out the three-dimensional distribution of galaxies in the Universe. The largest of these is the Euclid Consortium, with over 1200 members. That collaboration is responsible for the scientific program of ESA's Euclid mission, to be launched in 2020-2021 to study dark energy and dark matter in the universe.

This growth in astronomical research and research discoveries is also visible both in the number of PhDs conferred by universities and the number of postdoctoral fellows in the field. In the US, universities now confer around 150 PhDs per year in astronomy, a 50% growth in a decade from an average around 100 in mid 2000s[1]. Astronomy departments in US universities that award PhD degrees enrolled close to 1100 graduate students in 2013[2]. At the undergraduate level, astronomy Bachelor's degrees have nearly doubled from below 200 to nearly 400, again in the last decade[3]. While these are numbers from astronomy departments, a similar growth in physics PhDs related to cosmology or related subjects also add to number of post-PhD junior scientists searching for postdoctoral employment in astronomy and astrophysics in the US every year. Finally, the growth in astronomy PhDs from Europe and elsewhere mimics that of US. Over a 5-year interval from 2008 to 2012, the number of Physics PhD degrees awarded by UK universities increased by 25% from the low 600s to the high 700s[4] per year, despite an overall decrease in research funds in UK over the same period. The growth is also visible in the overall structure of large science collaborations. Close to 500 of the 1200 members of the Euclid Consortium, mentioned above, are junior scientists, either graduate students or postdoctoral fellows that will be looking for permanent positions in the field in a few years.

---

[1] http://www.aip.org/sites/default/files/statistics/undergrad/enrolldegrees-a-12.3.pdf
[2] http://www.aip.org/sites/default/files/statistics/rosters/astrost13.pdf
[3] http://www.aip.org/sites/default/files/statistics/undergrad/enroll%26degree-a-10.pdf

Unfortunately, despite the increase in the number of PhD degrees awarded per year in astronomy, the number of permanent scientific and academic positions has not kept up at the same rate. In the US, the average number of permanent positions advertised in the American Astronomical Society's Job Register over the last five years is at the level of 60-75 per year. This number is also consistent with the number of entries listed on the "astro rumor mill," an open-sourced website the astronomy community uses to gauge the academic job market, along with applicants that are shortlisted and eventually hired for those positions[5]. The best example in the growth of permanent positions in astronomy came from the UK during 2000-2010 with a 100% increase in the total number of professor positions from about 100 to the low 200s and a 50% increase in lectureships positions[6]. It will be interesting to see if this rapid growth has continued into this decade but indications are that this was a unique situation resulting from new schemes introduced in UK for appointment of professors on merit and the UK Research Excellence Framework (REF) assessment exercise. It is also becoming increasingly common to hear of single job advertisements for a junior faculty position in the US or Europe that has attracted 200 or more applicants. It is also now more common than ever before for US- or European-trained PhDs to seek permanent academic positions abroad, especially in countries such as Chile, South Africa and China with a growing infrastructure for astronomical research. Even in such countries we find that the number of applicants for junior faculty positions have doubled over the last few years. From all these indications it is clear that there is a disparity between the availability of permanent jobs and the production rate of PhDs on our campuses worldwide.

There are many reasons for this disconnect. One of them involves the nature of astronomy as a research field. While many fields largely rely on government research grants for growth, a typical astronomer or an astronomy group within a university does not require significant financial resources in the form of personal laboratory equipment to be productive in research, unlike biologists, chemists and some physicists. The significant costs exist at the level of national facilities, such as optical and radio observatories spread around the globe, and space telescopes, all of which are supplying immense datasets to the field. A few nights of observing time on a large observatory or a few orbits of the Hubble Space Telescope, equivalent to a few hours, are enough to keep a small team engaged in research over a year. The role of a typical astronomy group, even within large collaborative structures, then mainly involves the analysis and interpretation of those data. For that the main expenses are in labor. Since labor costs are lower at the level of students and postdoctoral fellows, small grants support the bulk of the astronomy research, funding primarily students and postdocs in the field. It is also easy for academic departments to increase the scientific output by modest payments for data access by joining large collaborations, followed by increases in the workforce at the junior level. This has resulted in a situation where these junior scientists, especially postdoctoral fellows, are continuously in the training mode from one position after another.

---

[4] http://www.ref.ac.uk/media/ref/results/AverageProfile_9_Physics.pdf
[5] Astro rumor mill http://www.astrobetter.com/wiki/Rumor+Mill
[6] https://www.ras.org.uk/images/stories/Publications/demography_of_astronomy.pdf

In the US, another reason for the aforementioned disparity is that the number of retirements by the baby boomer generation of faculty has not kept up with some of the early predictions[7]. Without a mandatory retirement many senior faculty members do not feel obliged to consider it. In the US, Fidelity[8] found that 74% of the faculty aged 49-67 years plan to either delay retirement past the nominal retirement age of 65 or not retire at all. The desire to stay productive and busy past retirement age, rather than financial considerations, was the primary reason given for delaying or avoiding retirement. This leads to a further increase in the pressure on the job market.

The biggest disconnect between the available permanent research employment opportunities in the field and the supply of PhDs happens at the university level. It is hard for universities to limit the number of students they enroll or to limit the number of students to whom they confer PhD degrees, once those students have enrolled in their programs. The current academic system focuses on research results in peer-reviewed journals within one chosen, usually narrowly defined, area of research that takes on average about four (in Europe) to six (in the US) years to complete. In a balanced environment, such a PhD would contain high quality research leading to a postdoctoral position for the student whom, after a nominal period of postdoctoral training, will fill a permanent academic or research position. This model worked perfectly in a bygone era when PhD meant a reasonable opportunity for a faculty or academic research position. With the growing disconnect between the number of PhDs and the available academic jobs these days, the previous academic model where permanent academic or research employment was the default for every PhD can no longer be the single option.

Despite this, astronomy as a research field still continues to function as if the balanced model is still viable. *New Worlds, New Horizons in Astronomy and Astrophysics* published by the US National Research Council (NRC) in 2010 only had recommendations on increasing the employment opportunities within astronomy. The report summarizes the 2010 Decadal Survey in astronomy, a community-organized exercise every decade that prioritizes research programs by funding agencies and provides recommendations on a variety of issues, including the status of employment and diversity in the field. It is easy to ignore the issue of permanent employment for our PhDs, or to pretend that it will automatically solve itself when the considerable surplus of postdoctoral fellows eventually leave the field, departing postdocs are simply replaced by a new set of PhDs, and such a hands-off approach does not solve any of the problems. It is also not healthy for astronomy as a research field when many junior scientists face an uncertain future.

---

[7] https://www.insidehighered.com/news/2013/06/17/data-suggest-baby-boomer-faculty-are-putting-retirement
[8] http://www.fidelity.com/inside-fidelity/employer-services/three-fourths-of-higher-education-baby-boomer-faculty-members

In order to change the culture within astronomy and move away from business as usual, the community, especially faculty and university administrators, must first acknowledge the existence of a significant academic employment issue despite the large number of PhDs that are awarded. Second, the field must move away from the established tradition in which faculty members solely train students for research with the immediate goal of finding a postdoctoral position within the field after PhD completion. Third, students who enter the PhD programs in astronomy should not pursue scientific research with the only goal of finding permanent academic or research employment. The situation is such that academic employment is no longer the first option or just one of many equal options. The PhD should be pursued with employment outside of academia as the main priority, with training in research methods and techniques as the top desirable aspect of that intended career.

The existing academic structure of training students solely for academic employment is forced partly by the research grants, internal campus promotions, and scientific societies that place a high value on placing students in successful careers in research. The structure discounts successes in placing students in careers outside of academic research. In our new proposed model the PhD production has to be balanced by the available jobs in both industry and academia. To successfully move away from the existing model, we also need to produce PhD-trained students that are competitive applicants for jobs outside of the academic structure. Thus, our programs must confer to the requirements of an industry that has jobs currently available, such as tech, finance, or secondary education. This is challenging since academic programs are slow to adapt to changes and demands outside of universities, while new industries go through cycles in decade-scale timescales. In the present cycle, astronomy students are best situated for new developments in the tech industry. Our latest research with "big data" astronomy projects explores the boundaries of software engineering, statistics, machine learning, data mining, and database management, among many other novel developments. If academic programs can be structured meaningfully such that students are given the chance to explore the applications of these ideas outside of pure astronomical research, they can be competitive in the job market both in and out of academic research.

We are a group of astronomers from the largest survey program that the Hubble Space Telescope has conducted (CANDELS). Like many large astronomy survey teams of its kind, our collaboration involves several hundred scientists and students spread across the globe. Amongst our team, more than 60% are junior scientists who will face these employment challenges in the near future. We outline here a set of recommendations that are aimed towards changing the current employment culture for astronomers, with the ultimate goal of expanding such changes in the future to create a bigger impact in the field. Our recommendations are mostly for changes within astronomy as a research field, but can be easily modified and adopted by other fields where similar situations now exist. We separate these recommendations and policy changes at all levels; from large research collaborations, to universities, and to governmental funding organizations. All these levels of the research infrastructure have an equal stake in the status of the employment within the field and should pursue a variety of changes, including the ones that we outline, to mitigate the growing disconnect.

*1. Within large collaborations:* CANDELS has an active junior scientist collaboration, organized and led by junior scientists. They meet independently to discuss activities within the research team and to discuss employment opportunities. At collaboration meetings we have organize opportunities for junior scientists as groups to meet with former researchers who have made successful transitions to careers industry, away from universities and national labs. This is a policy that can be implemented at the level of funding agencies, or by organizations such as NASA and ESA, when funding a large research endeavor in astronomy or cosmology. The PIs or leaders of large research collaborations must also take an active role to monitor the employment status of junior members of the team and to make publicly available such statistics, just as statistics on research progress is made available and reported to national agencies as part of renewal grants for the research effort. Given that astronomy is, or will soon be, structured into large collaborations, big changes in the field as a whole are likely to come through restructuring of priorities individually within such research collaborations. We encourage open communication within such teams about employment opportunities outside of academic research. This will help overcome the existing close-minded nature within departments and universities on the employment situation and encourages graduate students who are members of these teams to prepare for employment outside of academia. Meetings with ex-scientists from the field will also allow students to see and learn how their skills could be useful for certain industries.

*2. Within universities:* The current academic system must slowly change internally so that placement of PhD students in any desirable job is equally valued as placing a student in an academic postdoctoral position. This is a policy that is easy to implement within astronomy/physics PhD programs with small changes to the curriculum, instead of relying on outside entities, such as Insight Data Science Fellows Program, to train PhD scientists for industry jobs. Astronomy students are desirable for many tech and finance industry jobs given their ability to analyze large datasets, delve deeply into statistics, and pursue interdisciplinary studies easily, among other qualities. Employers are also changing to the needs of a culture that values independent research and thinking, a desirable objective that is expressed by many students when stating why they wish to remain in academic jobs despite the low salaries and large uncertainties on future employment. Such freedom is also acknowledged by some of the large companies like Google that allow their employees to use a fraction of their time to pursue independent projects. Finally, universities must change in ways to meet the requirements of these industries so that about half of the PhDs conferred are aimed directly for post-PhD employment outside of the academia. This could be facilitated through internships in the industry as a replacement for one fewer research publication or a set of academic course credits to satisfy the PhD requirements.

    3. *National Societies:* National societies in astronomy, such as AAS or RAS, should take an active role in changing the culture within the field. We suggest the following activities, which are each aimed at its own demographic in the academic hierarchy. It is now common to organize workshops aimed towards junior and new faculty, with primary focus on the career development and training for effective teaching. These workshops could also include open discussions on the employment situation within the field, and recommendations for mentoring in light of this reality. As an example, junior faculty, who are more likely to accept and introduce new changes than more senior and established faculty, should be encouraged to train their students on new software languages and tools that are different from those that they themselves may have used during graduate school. Although initially this may be a short-term sacrifice in science productivity, it is likely to facilitate growth in the astronomical community as people learn innovative techniques being adopted in industry. The basic message to junior faculty at these workshops should be that they will not be considered failures but will be successful if they train and prepare PhD students for employment outside of the academic research structure.  Second, the national societies should provide incentives that appeal to established senior members. This could take the form of awards or prizes for mentorship that bridges the gap between astronomy and industry, or grants with part of it paying student stipends while they do industry internships. Third, national societies should maintain a networking database of past and current members who work outside of academia. The responsibility should fall to the national societies, as they have the means to semi-annually solicit all current and former members to submit their current employment status. Such a resource should include the job title, organization and contact information of members who have found gainful

employment outside of academia. This resource would facilitate communication between current and former PhD students, so the current students can easily contact people who currently live in their desired area and work in their desired field to talk about their job and preparation. This form of mentorship, the simple exposure to people with the same educational background and their successes, is arguably one of the most lacking. Such networking is also indispensable in a career transition

4. *At the level of funding agencies:* We encourage funding agencies, such as the Astronomy Division within NSF or NASA/ESA, to implement policies that support student training that is not focused only for academic research careers. For example, the NSF in the US requires a one-page statement from PIs on the mentoring of junior researchers, such as postdoctoral fellows, in their research groups as part of grant applications. While not explicitly required by the NSF, such statements are almost always biased towards scenarios where postdoctoral fellows are assumed to become a future faculty member. Furthermore, review panels continue to weigh heavily on the placement of researchers in highly desirable faculty positions. A policy must be adopted that mentoring statements consider employment issues more broadly and allow PIs to mentor students and postdoctoral for jobs outside of academia. We encourage review panels, made up of our peers, to acknowledge the difficulties faced in the academic market and to avoid discounting PIs who are actively involved in efforts to help junior scientists transition from academic research to jobs in desirable industries. We also encourage funding agencies to require a networking plan to facilitate communication between junior scientists in large collaborations and former academics that have made a successful transition away from academic research. This could be part of the employment monitoring in our first recommendation.

Astronomy produces PhD-trained individuals who have strong technical skills and are exceptional problem solvers: everyone in the field should be invested in placing our graduates, who contribute the bulk of research activity, into the top positions they deserve, whether this is in academia or in industry.

*Asantha Cooray is a Professor in Physics at the University of California, Irvine. Boris Häußler is a staff astronomer at ESO/Chile. Alexandra Abate and Christina Williams are postdoctoral fellows at University of Arizona, and Jonathan Trump is a postdoctoral fellow at Penn State University.*